\documentclass[twocolumn,aps,amssymb,superscriptaddress]{revtex4}
\usepackage{graphicx}

\begin{document}
\title{Far-infrared and submillimeter-wave conductivity in electron-doped cuprate La$_{2-x}$Ce$_{x}$CuO$_{4}$ }
\author{A. Pimenov}
\affiliation{Experimentalphysik V, EKM, Universit\"{a}t Augsburg, 86135 Augsburg,
Germany}
\author{A. V. Pronin}
\affiliation{Experimentalphysik V, EKM, Universit\"{a}t Augsburg, 86135 Augsburg,
Germany}%
\affiliation{General Physics Institute of the Russian Acad. of Sciences, 119991 Moscow,
Russia}
\author{A. Loidl}
\affiliation{Experimentalphysik V, EKM, Universit\"{a}t Augsburg, 86135 Augsburg,
Germany} %
\author{A. Tsukada}
\affiliation{NTT Basic Research Laboratories, 3-1, Morinosato-Wakamiya, Atsugi-shi, Kanagawa 243-0198, Japan}%
\author{M. Naito}
\affiliation{NTT Basic Research Laboratories, 3-1, Morinosato-Wakamiya, Atsugi-shi, Kanagawa 243-0198, Japan}%

\date{\today}

\begin{abstract}
We performed far-infrared and submillimeter-wave conductivity experiments in the
electron-doped cuprate La$_{2-x}$Ce$_{x}$CuO$_{4}$ with $x=0.081$ (underdoped regime,
$T{\rm _c}=25$\,K). The onset of the absorption in the superconducting state is gradual
in frequency and is inconsistent with the isotropic s-wave gap. Instead, a narrow
quasiparticle peak is observed at zero frequency and a second peak at finite frequencies,
clear fingerprints of the conductivity in a d-wave superconductor. A far-infrared
conductivity peak can be attributed to $4 \Delta_0$, or to $2\Delta_0+\Delta_{spin}$,
where $\Delta_{spin}$ is the resonance frequency of the spin-fluctuations. The infrared
conductivity as well as the suppression of the quasiparticle scattering rate below $T{\rm
_c}$ are qualitatively similar to the results in the hole-doped cuprates.

%However, important differences from these behavior can be observed, such as a reduced
%frequency of the conductivity peak, which indicate different superconducting mechanism in
%La$_{2-x}$Ce$_{x}$CuO$_{4}$.
\end{abstract}


\maketitle

\section{Introduction}
The interest in the physical properties of electron-doped superconductors \cite{tokura}
has revived recently concerning the symmetry of the superconducting order parameter.
Earlier results in these compounds on penetration depth \cite{wu}, Raman \cite{stadlober}
and tunneling spectroscopies \cite{alff} were explained in terms of conventional (s-wave)
symmetry of the superconducting order parameter. However, later experiments, including
half-flux effect \cite{tsuei}, penetration-depth measurements \cite{prozorov}, and
photoemission \cite {sato} provided strong evidences for d-wave type symmetry. This
contradiction can possibly be resolved on the basis of recent microwave experiments
\cite{lemberger} and point-contact spectroscopy \cite{biswas}, which suggest changes of
the gap symmetry as a function of doping.

Electron doping of the high-$T{\rm _c}$ cuprates can be achieved by substituting
Ce$^{4+}$ into Ln$_{2}$CuO$_4$ with Ln = Pr, Nd, Sm, and Eu \cite{tokura,naito}. Among
these family Nd$_{2-x}$Ce$_x$CuO$_4$ is the earliest known and best studied compound
\cite{fournier}. The highest transition temperatures for electron-doped cuprates ($T{\rm
_c}=30\,$K) can be achieved in La$_{2-x}$Ce$_x$CuO$_4$ (LCCO) \cite{naito}. The
temperature of the superconducting transition in LCCO is close to $T_{\rm c}=39\,$K in a
recently-discovered superconductor MgB$_2$ \cite{nagamatsu}, which is believed to be of
BCS-type \cite{dahm,advances}. In order to obtain valuable information about the gap
symmetry in LCCO, a direct comparison of the physical properties of these compounds can
be carried out. Such analysis is provided by the low-frequency electrodynamics which
directly visualizes many important features of superconductors as energy gap
\cite{nb,mgb2ir} or quasiparticle scattering rate \cite{ybco}.

In this paper we present the far-infrared and submillimeter-wave conductivity of
underdoped ($x=0.081$, $T{\rm _c}=25\,$K) LCCO thin film. In order to obtain the complex
conductivity above and below the superconducting gap energy, two different experimental
methods have been applied using the same sample. For frequencies below 40\,cm$^{-1}$ the
complex conductivity has been obtained by the submillimeter transmission spectroscopy. At
higher frequencies the reflectance was measured using standard far-infrared techniques
and the conductivity has been obtained via a Kramers-Kronig analysis of the spectra.

\section{Experimental details}

High quality La$_{2-x}$Ce$_{x}$CuO$_{4}$ film with $x=0.081 \pm 0.01$ (underdoped
regime), were deposited by molecular-beam epitaxy \cite{naito} on transparent (001)
SrLaAlO$_{4}$ substrates $10 \times 10 \times 0.5$ mm$^3$ in size.  The thickness of the
present film was 140 nm and the film revealed a sharp transition into the superconducting
state ($\Delta T_{c} < 1$\,K) at $T_{\rm c}=25\,$K. Lower transition temperature compared
to the optimal doping ($x\simeq 0.11$, $T_{\rm c}=30\,$K \cite{naito}) confirms the
underdoped character of the sample.

The transmission experiments for frequencies $5$\,cm$^{-1}<\nu <40$\,cm$^{-1}$ were
carried out in a Mach-Zehnder interferometer arrangement \cite{volkov} which allows both,
the measurements of the transmittance and the phase shift of a film on a substrate. The
properties of the blank substrate were determined in a separate experiment. Utilizing the
Fresnel optical formulas for the complex transmission coefficient of the substrate-film
system, the absolute values of the complex conductivity $\sigma ^{*}=\sigma _{1}+i\sigma
_{2}$ were determined directly from the measured spectra. In the frequency range $40<\nu
<4000$\,cm$^{-1}$ reflectance measurements were performed using a Bruker IFS-113v
Fourier-transform spectrometer. In addition, the reflectance for frequencies $5<\nu
<40$\,cm$^{-1}$ was calculated from the complex conductivity data of the same samples,
which was obtained by the submillimeter transmission. This substantially improves the
quality of the subsequent Kramers-Kronig analysis of the reflectance and therefore the
reliability of the data especially at low frequencies. A similar technique has been
applied recently to the films of newly discovered MgB$_2$ \cite{mgb2ir,mgb2}, leading to
the observation of a superconducting absorption edge. Reference \cite{mgb2ir} gives
further details of the experimental set-up.

The reflectance of a thin metallic film on a dielectric substrate can be obtained from
the Maxwell equations \cite{heavens}:
\begin{equation}
r=\frac{r_{0f}+r_{fs}\exp ({4\pi }in_{f}d{ /\lambda )}}{1+r_{0f}r_{fs}\exp ({4\pi
}in_{f}d{/\lambda )}}\quad , \label{erefl}
\end{equation}
with $r_{0f}=(1-n_f)/(1+n_f)$ and $r_{fs}=(n_f-n_s)/(n_f+n_s)$ being the Fresnel
reflection coefficients at the air-film $(r_{0f})$ and film-substrate $(r_{fs})$
interface. Here $n_f = (i\sigma^* /\varepsilon_0 \omega^*)^{1/2}$ and $n_s$ are the
complex refractive indices of the film and substrate, respectively, $\lambda $ is the
radiation wavelength, $d$ is the film thickness, $\omega= 2\pi \nu$ is the angular
frequency, $\sigma^*=\sigma_1+i\sigma_2$ is the complex conductivity of the film, and
$\varepsilon_0$ is the permittivity of free space. Eq.\ (\ref{erefl}) is written
neglecting the multiple reflections from the opposite sides of the substrate.

If the film thickness is smaller than the penetration depth ($\left| n_{f}\right| d \ll
\lambda $) and if $\left| n_{f}\right| \gg \left| n_{s}\right| $, Eq. (\ref{erefl}) can
be simplified to :
\begin{equation}
r \approx \frac{1-\sigma ^{*}dZ_{0}-n_{s}}{1+\sigma ^{*}dZ_{0}+n_{s}} \label{erefl1} ,
\end{equation}
where $Z_{0}=\sqrt{\mu_0/\varepsilon_0} \simeq 377\,\Omega$ is the impedance of free
space. Eq. (\ref{erefl1}) provides a good approximation of the reflectance at
submillimeter frequencies. For higher frequencies Eq. (\ref{erefl}) has to be used.

\begin{figure}[]
\centering
\includegraphics[width=8cm,clip]{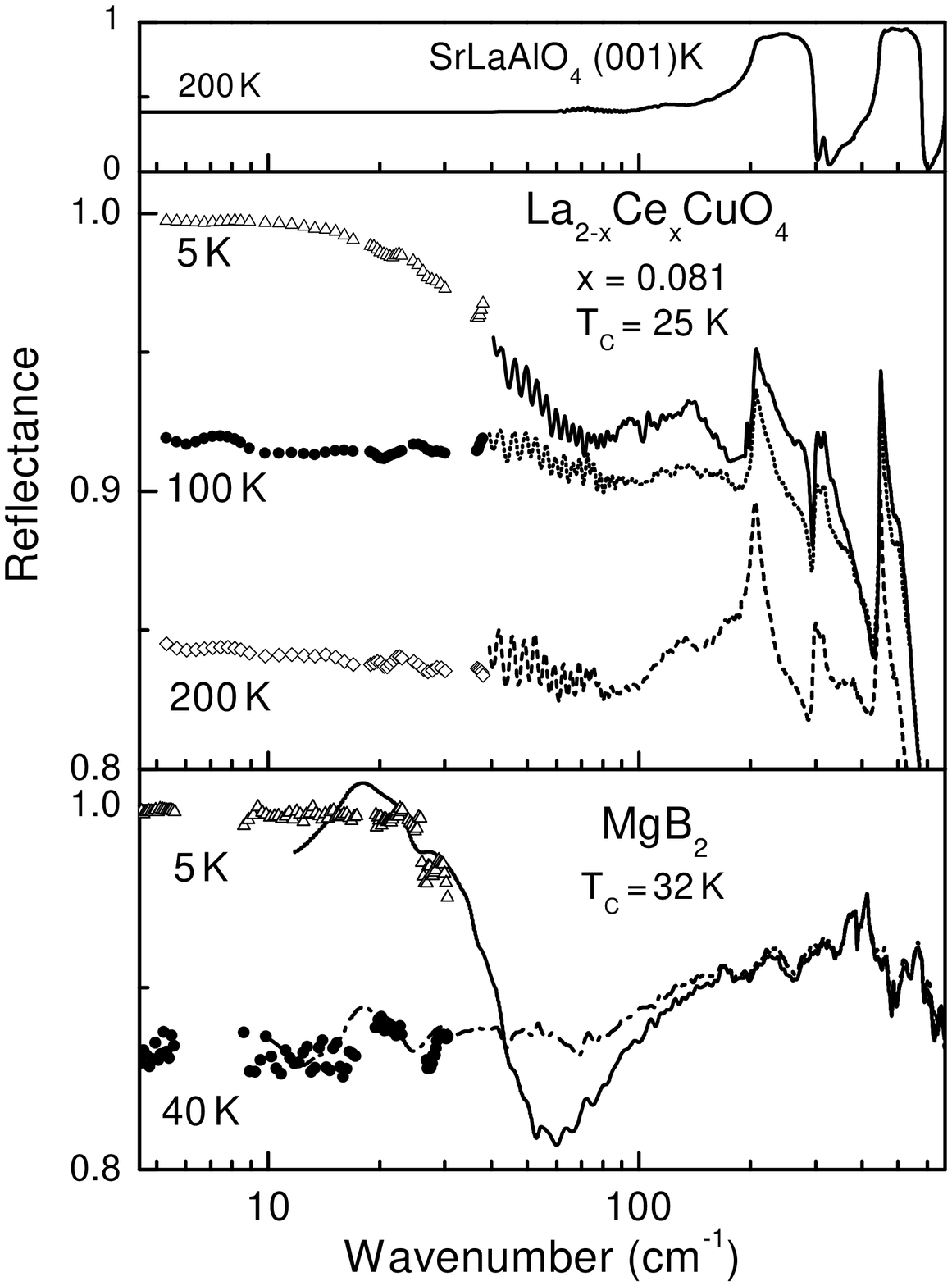}
%\vspace{0.2cm}
\caption{Far-infrared reflectance of underdoped LCCO film at different temperatures
(middle panel). Lines - directly measured data, symbols - reflectance as calculated from
the complex conductivity measured by submillimeter transmission technique. Upper panel
shows the reflectance of the blank substrate. Lower panel shows the reflectance of
MgB$_2$ film \cite{mgb2ir,advances} for comparison.} \label{frefl}
\end{figure}

\section{Results and Discussion}

The middle panel of Fig.\ \ref{frefl} shows the far-infrared reflectance of the
underdoped LCCO film at different temperatures. A common feature of all spectra is a
sharp structure above 200 cm$^{-1}$ which is due to the phonons of the substrate (upper
panel). The influence of the substrate is
 reduced substantially by
 calculating the complex conductivity via Eq. (\ref{erefl}) but cannot be
fully removed. At low frequencies and in the normally-conducting state the reflectance is
approximately frequency independent. This is in agreement with Eq. (\ref{erefl1}) for a
metal at low frequencies with $\sigma^* \simeq \sigma_1$. In the superconducting state,
the low-frequency reflectance of the LCCO film becomes frequency dependent and follows
approximately $1-|r|^2 \propto \nu^2$. This behavior follows directly from Eq.
(\ref{erefl1}): in the superconducting state the complex conductivity can be approximated
by $\sigma^* \simeq i\sigma_2\propto i/\nu$, which leads to $1-|r|^2 \propto \nu^2$.
Comparing the reflectance of LCCO in the superconducting state (Fig. \ref{frefl}) with
the spectra of a s-wave superconductor with comparable transition temperature, e.g. with
the reflectance spectra of MgB$_2$ (lower panel) \cite{mgb2ir,advances}, significant
differences can be observed. In the spectra of MgB$_2$ the s-wave symmetry of the
superconducting order parameter leads to a sharp "knee" in the reflectance around $h \nu
\simeq 2\Delta_0$, which is in contrast to the spectra of LCCO where only a gradual
decrease is observed. Thus, already the analysis of the reflectance spectra reveals a
first indication of an unconventional gap-symmetry in LCCO.

\begin{figure}[]
\centering
\includegraphics[width=7cm,clip]{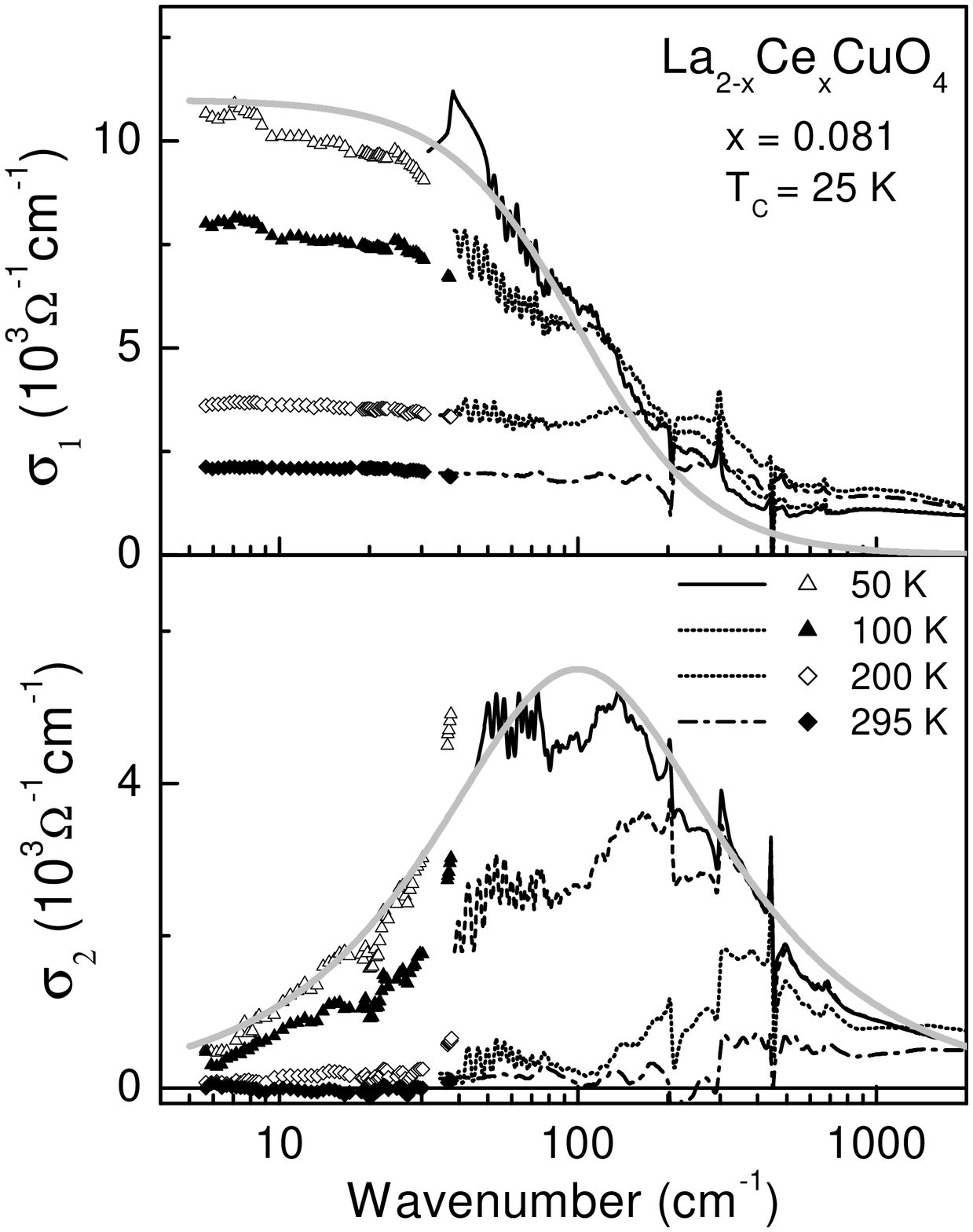}
%\vspace{0.2cm}
\caption{Far-infrared complex conductivity of underdoped LCCO film above $T_{\rm c}$.
Upper panel - $\sigma_1$, lower panel - $\sigma_2$. Lines represent the conductivity
obtained from the infrared reflectance, symbols - the conductivity as measured directly
by the submillimeter transmission technique. Thick gray line shows the prediction of the
Drude model with $1/ 2 \pi \tau = 100$\,cm$^{-1}$.} \label{fsign}
\end{figure}

Figure \ref{fsign} shows the far-infrared conductivity of the underdoped LCCO film in the
normally-conducting state. The results above 40\,cm$^{-1}$ were obtained applying the
Kramers-Kronig analysis to the reflectance data and  solving Eq. (\ref{erefl}). Below
40\,cm$^{-1}$ the complex conductivity was calculated directly from the transmittance and
phase shift. We recall that the resonance-like structures between 200 and 700\,cm$^{-1}$
are due to the residual influence of the substrate. In this frequency range only the
overall frequency dependence of the conductivity is significant. The far-infrared
conductivity in the normally-conducting state can well be described by the Drude model
with a frequency-independent scattering rate $1/ \tau$. At low frequencies, $\sigma_1$ is
frequency-independent and $\sigma_2$ increases approximately linearly with frequency. For
frequencies close to the value of the scattering rate, $\sigma_1$ starts to decrease and
$\sigma_2$ shows a maximum, $\nu_{max}\simeq 1/ 2 \pi\tau$. The gray solid line in Fig.
\ref{fsign} provides a good description of the conductivity at $T=50\,$K and demonstrates
the validity of the Drude model for LCCO. Substantial deviations between the data and the
model can be observed above 200\,cm$^{-1}$. These deviations can be described assuming a
frequency dependence of the quasiparticle scattering, which agrees well with the infrared
experiments in (NdCe)$_2$CuO$_4$ by Homes \textit{et al.} \cite{homes} and Singley
\textit{et al.} \cite{singley}.

The infrared spectra of high-$T_{\rm c}$ cuprates and of highly anisotropic materials
often show a finite-frequency peak in $\sigma_1 (\omega)$. In addition to well-known
mechanisms to explain the origin of this peak, like localization of the charge carriers
\cite{timusk}, a new possibility has been proposed recently \cite{tilt}. The conductivity
peak at finite frequencies results from a small ($\sim 1^{\circ}$) tilt of the sample
surface from the ideal c-axis orientation. In present experiment care has been taken to
avoid such tilt effects, resulting in a peak-free real part of the infrared conductivity
in the normal state.

\begin{figure}[]
\centering
\includegraphics[width=7cm,clip]{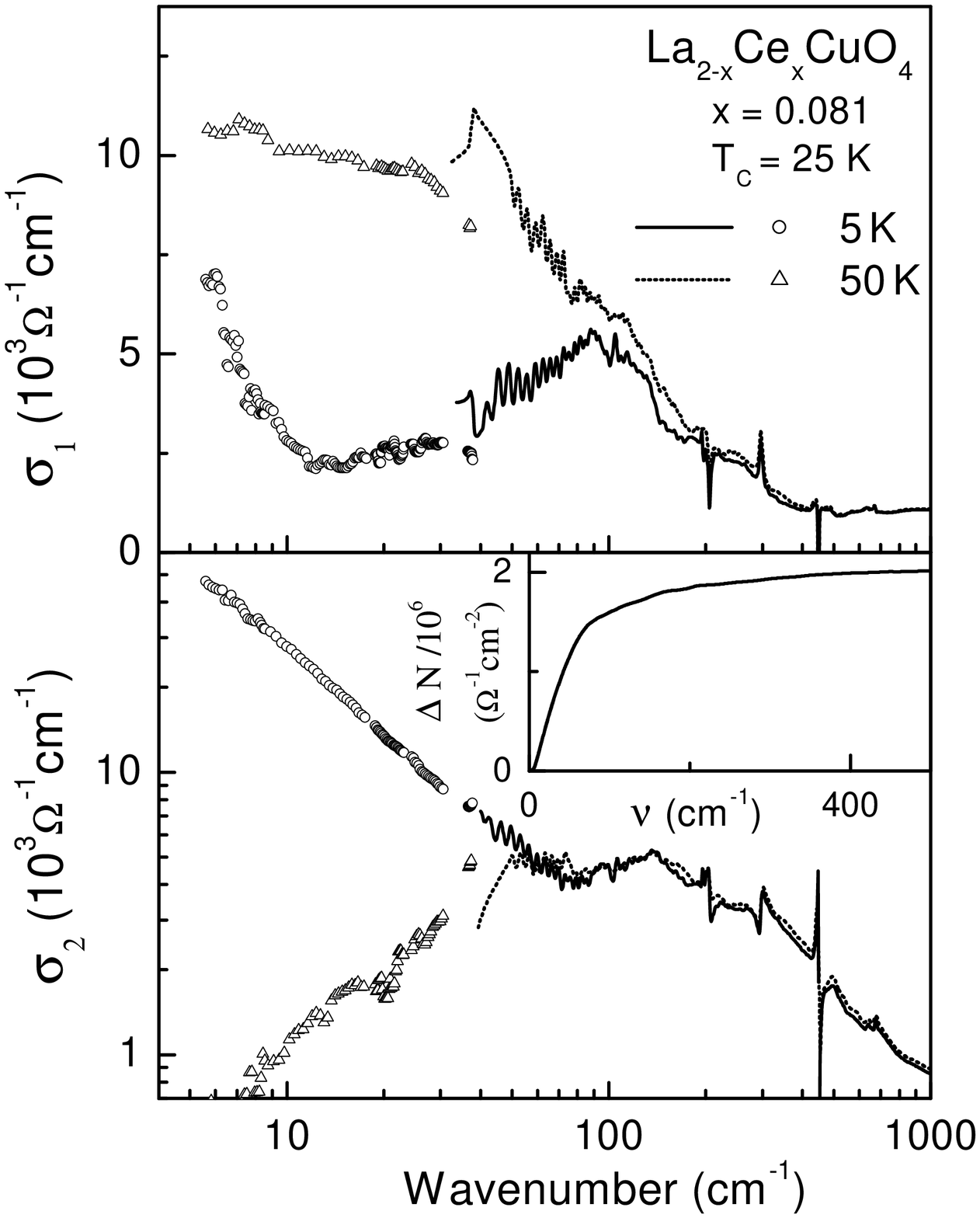}
%\vspace{0.2cm}
\caption{Far-infrared complex conductivity of underdoped LCCO film above
and below $T{\rm _c}=25\,$K. Lines represent the conductivity obtained from the infrared
reflectance, symbols - the conductivity as measured directly by the submillimeter
transmission technique. Inset shows the frequency dependence of the missing spectral
weight, Eq. (\ref{eqsum}).} \label{fsigs}
\end{figure}

The infrared conductivity changes dramatically upon entering the superconducting state.
These data are represented in Fig. \ref{fsigs}. The real part of the conductivity is
strongly suppressed below 100\,cm$^{-1}$ (upper panel). Contrary to the conductivity of
s-wave superconductors \cite{mgb2ir,palmer}, no clear onset of $\sigma_1(\nu)$ can be
seen in the far-infrared range, which again indicates an unconventional character of the
energy gap. Instead, $\sigma_1(\nu)$ clearly reveals two contributions, a Drude-like
quasiparticle peak at zero frequency and a finite-frequency peak close to 100\,cm$^{-1}$.
A similar maximum at correspondingly higher frequencies has been observed in infrared
experiments on hole-doped cuprates \cite{basov}. Assuming a spin-fluctuation scenario of
superconductivity, the frequency of the conductivity peak can be ascribed to the
quadrupled frequency of the superconducting gap $4 \Delta_0$ \cite{quinlan}. Compared to
an s-wave superconductor with the conductivity onset at $2 \Delta_0$, an additional shift
by $2 \Delta_0$ arises due to a four-particle final state. The d-wave pairing in
connection with a spin-fluctuation mechanism may lead to another characteristic energy
scale of the optical conductivity \cite{abanov}. In that case the residual attraction in
a d-wave superconductor binds a particle and hole in a spin exciton at an energy
$\Delta_{spin}$. As a result the characteristic feature in the conductivity is shifted to
$2\Delta_0+\Delta_{spin}$ \cite{abanov}. These mechanisms possibly explain the origin of
the conductivity peak shown in Fig.\ \ref{fsigs}. Although the optical spectroscopy is
not sensitive to a sign change of the superconducting order parameter, the conductivity
data in Fig.\ \ref{fsigs} provide strong experimental evidence for a highly anisotropic
(and, possibly, d-wave) energy gap in underdoped LCCO: according to d-wave model
calculations including spin-fluctuation scattering, a gap with nodes gives rise to a
residual Drude-like peak \cite{hirschfeld}, and a peak at finite frequencies resulting
from inelastic scattering processes \cite{quinlan,abanov}.

%However, an important quantitative difference to the properties of YBaCuO is the
%relatively low frequency of the conductivity peak compared to the values of transition
%temperature: $h\nu_{max}/kT{\rm _c}\simeq 13$ for YBaCuO \cite{basov}, while
%$h\nu_{max}/kT{\rm _c}\simeq 5.2$ for LCCO. These ratios differ by more than factor of 2,
%which  indicates significant differences in the superconducting mechanism in these
%compounds.

Due to the suppression of the conductivity at  far-infrared frequencies, substantial
spectral weight is removed from this frequency range and is transferred to the
delta-function at zero frequency (superconducting condensate). This transfer leads to the
dramatic increase of $\sigma_2(\nu)$ at low frequencies,  $\sigma_2(\nu) = A / \omega$.
The spectral weight of the superconducting condensate can easily be obtained as the
pre-factor of this proportionality $A=n_s e^2/m=1.7\cdot10^6 \Omega^{-1}cm^{-2}$. The
missing spectral weight (partial sum rule \cite{ir}) can be calculated  by direct
integration
\begin{equation}
\Delta N (\omega)=\frac{2}{\pi}\int\limits_{0}^{\omega}[\sigma_{1,n}-\sigma
_{1,s}](\omega )d\omega \quad . \label{eqsum}
\end{equation}
The result is shown in the inset of Fig. \ref{fsigs}. $A=\Delta N$ indicates the
conservation of the spectral weight. The change of spectral weight, $\Delta N$, saturates
only around $\nu\sim 400\,$cm$^{-1}$, i.e. for frequencies well above the characteristic
gap frequency. In LCCO the far-infrared saturation gives $\Delta N = 1.9\cdot10^6
\Omega^{-1}cm^{-2}$ and is $\sim 20\%$ higher than the measured weight of the condensate.
This difference probably indicates that some amount of the spectral weight remains in the
superconducting state at frequencies below the range of the present experiment.

At frequencies below 10\,cm$^{-1}$ and in the superconducting state, $\sigma_1$ shows a
wing of the low-frequency excitations (upper panel of Fig. \ref{fsigs}) which probably
corresponds to a Drude-like quasiparticle peak \cite{quinlan,hirschfeld}. The rate of the
quasiparticle scattering is strongly suppressed compared to the normal-state ($1/2 \pi
\tau \simeq 100\, $cm$^{-1}$ at $T=50\,$K) and attains values around 10\,cm$^{-1}$.

\begin{figure}[]
\centering
\includegraphics[width=6.5cm,clip]{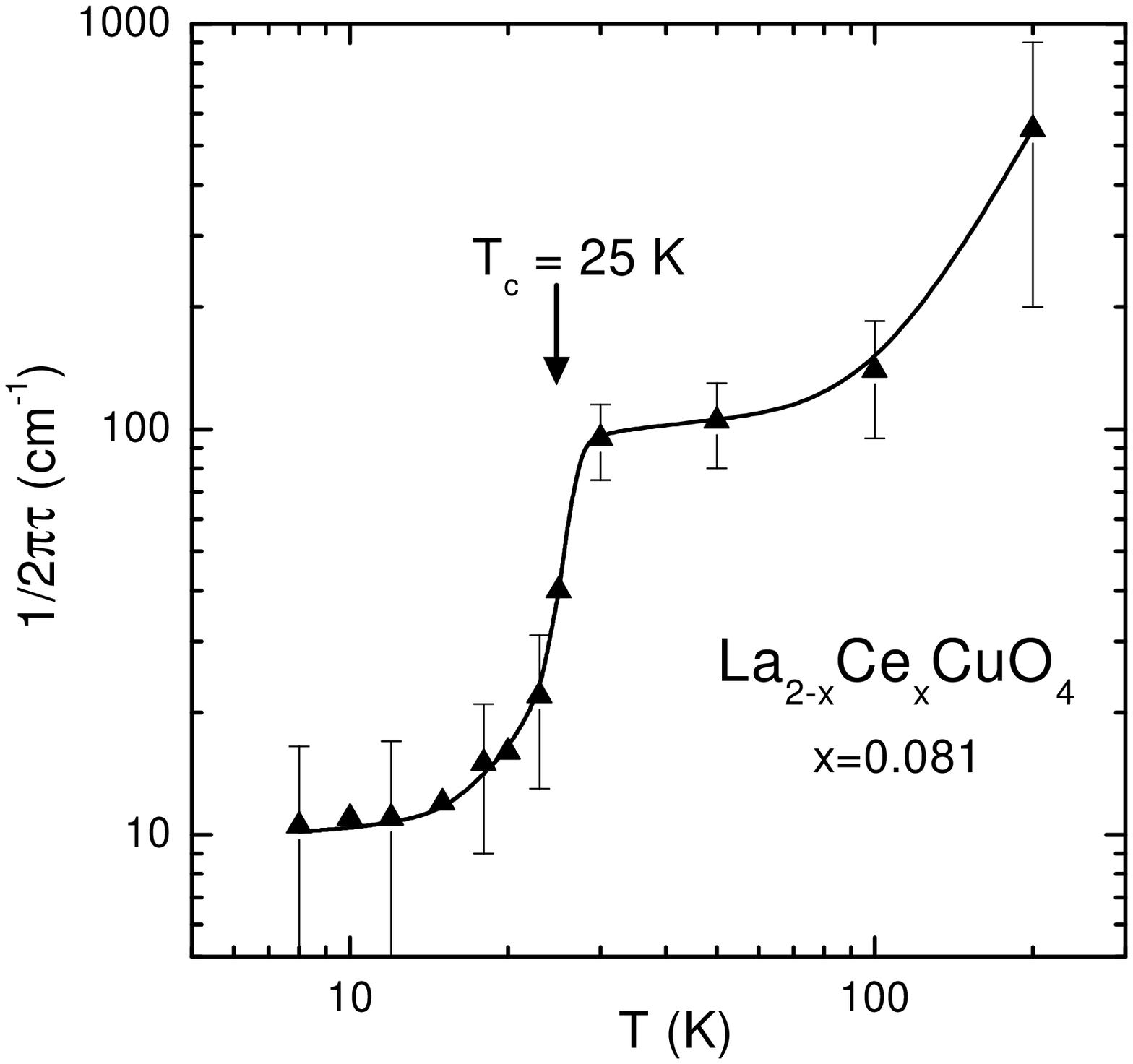}
%\vspace{0.2cm}
\caption{Temperature dependence of the effective quasiparticle scattering in LCCO. The
data have been obtained from the two-fluid analysis of the submillimeter-wave
conductivity, Eq. (\ref{eqdrude}).} \label{ftau}
\end{figure}

Fig. \ref{ftau} shows the temperature dependence of the quasiparticle scattering of LCCO.
The effective scattering rate has been obtained solely from the submillimeter-wave
conductivity using a two-fluid analysis \cite{ybco}
 \begin{equation}
\sigma^*(\omega)=  \frac{\varepsilon_0 \omega_p^2
\tau}{1-i\omega\tau}+A[\frac{\pi}{2}\delta(0)+\frac{i}{\omega}] \quad . \label{eqdrude}
\end{equation}
Here $\omega_p^2$, $\tau$, and $A$ represent  the plasma frequency, the scattering rate
of quasiparticles, and the spectral weight of the superconducting condensate. In this
equation the delta function $\delta(0)$ obviously does not influence the calculations at
finite frequencies and the parameter $A$ is obtained as low-frequency limit of
$\sigma_2\cdot \omega$.

The most prominent feature of Fig. \ref{ftau} is the suppression of the effective
scattering rate directly at $T{\rm _c}$. This is  similar to the temperature dependence
of the scattering rate in optimally-doped YBa$_2$Cu$_3$O$_{7-\delta}$, where a drop in
$1/\tau$ has been observed, e.g. using microwave resonator technique \cite{bonn} or
submillimeter transmission spectroscopy \cite{ybco}. However, in case of
YBa$_2$Cu$_3$O$_{7-\delta}$ the scattering rate revealed a linear temperature dependence
above $T{\rm _c}$, in contrast to LCCO where the scattering rate levels off for
temperatures below $\sim 100\,$K.

\section{Conclusions}

In conclusion, combining two experimental techniques we obtained the far-infrared
conductivity of underdoped La$_{2-x}$Ce$_{x}$CuO$_{4}$ in the frequency range above and
below the gap frequency. No characteristic onset of absorption is observed in the
superconducting state, which is inconsistent with the conventional BCS scenario. At low
temperatures a maximum of infrared conductivity is observed at frequencies close to
100\,cm$^{-1}$ which is qualitatively similar to the properties of the hole-doped
cuprates. The quasiparticle scattering rate is suppressed upon entering the
superconducting state. These results provide experimental evidence for a d-wave or highly
anisotropic s-wave gap in underdoped LCCO.

\section{Acknowledgements}

The stimulating discussion with P. J. Hirschfeld is gratefully acknowledged. This work
was supported by BMBF (13N6917/0 - EKM).

\end{document}